\documentclass{ifacconf}
\pdfoutput=1
 \usepackage{times}
  \usepackage[small,compact]{titlesec}
   
\usepackage{tipa}
\usepackage{amsmath}
\usepackage{amsfonts}
\usepackage{graphicx}      
\usepackage{adjustbox}
\usepackage[margin=2mm]{subfig}
\usepackage{amssymb}
\usepackage{pgfplots}
\usepackage{tikz}
\usepackage{natbib} 

\providecommand{\norm}[1]{\lVert#1\rVert}

\newcommand{\ie}{\mathsf{i.e.}}
\newcommand{\R}{\mathbb{R}}

\newcommand{\braces}[1]{\left({#1}\right)}
\newcommand{\nextline}{\nonumber \\&}

\newtheorem{defin}{Definition}

\newtheorem{Corr}{Corollary}
\newtheorem{Thrm}{Theorem}
\begin{document}

\begin{frontmatter}
\title{Consensus analysis of systems with time-varying interactions : An event-triggered approach} 
\author[First]{S. Arun Kumar} 
\author[First]{ N. R. Chowdhury} 
\author[First]{S. Srikant}
\author[Second,Third]{J. Raisch}

\address[First]{Department of Systems and Control Engineering, Indian Institute of Technology Bombay, Mumbai -400076, India. (e-mail:$\{\text{arunkumar92,~nilanjan,~srikant}\}$@sc.iitb.ac.in)}
\address[Second]{Fachgebiet Regelungssysteme, Technische Universit\textipa{\"a}t
Berlin, Berlin, 10587 Germany (e-mail: raisch@control.tu-berlin.de).}
\address[Third]{Systems and Control Theory Group, Max Planck Institute for Dynamics of Complex Technical Systems.}

\begin{abstract}
We present consensus analysis of systems with single integrator dynamics interacting via time-varying graphs under the event-triggered control paradigm. Event-triggered control sparsifies the control applied, thus reducing the control effort expended. Initially, we consider a multi-agent system with persistently exciting interactions and study the behaviour under the application of event-triggered control with two types of trigger functions- \textit{static} and \textit{dynamic} trigger. We show that while in the case of static trigger, the edge-states converge to a ball around the origin, the dynamic trigger function forces the states to reach consensus exponentially. Finally, we extend these results to a more general setting where we consider switching topologies. We show that similar results can be obtained for agents interacting via switching topologies and validate our results by means of simulations.
\end{abstract}
\end{frontmatter}
\section{Introduction}
A flock of birds, a swarm of bees, a school of fishes, a colony of ants -all display a wonderful coordination and complex patterns which have caught our attention since time immemorial. On closer observation we realize that these seemingly complex patterns emerge out of simple yet powerful, local rules on each agent of the group based on interactions with only the neighbours. When we attempt to algorithmize such a 'decentralized' behaviour, two types of questions can be posed- what would be the result of a particular decentralized control law or what decentralized control law would result in a desired global pattern or formation. Consensus is one of the most common and powerful 'global' behaviours  usually studied, primarily because it can be easily extended to other problems like formation, rendezvous, flocking \emph{etc}. \cite{vicsek1995novel} proposed an average-based decentralized control law for a multi-agent system based on only local information and observed that the agents attain consensus while \cite{jadbabaie2003coordination} gave a proof of the same assuming the agents interact via connected graphs that can switch at different time instants. Various results have been proposed by \cite{ren2005consensus}, \cite{ren2005second}, \cite{olfati2005consensus}, and others on consensus of a multi-agent system under directed or undirected graphs, with switching topologies and also with time delays.
\\Studying consensus behaviour for systems with switching or time-varying graphs is naturally of interest as in real-life scenarios it is not possible to assume that each agent has the same set of neighbours at all times. Time-varying graphs are useful when the information from each neighbour is assigned a weight proportional to the reliability of the information or the distance between them. \cite{martin2013continuous} proved consensus under the assumptions of \emph{persistent}-connectivity and cut-balance interactions. \cite{chowdhury2016persistence} obtained bounds on the rate of convergence for single-integrator and double integrator dynamics under persistent interactions.\\
With control laws being implemented on digital computers, developing discrete-time counterparts to continuous-time control laws is an eventuality. Among discrete control laws event-triggered control is preferred over time-triggered control as it comes into play only when an 'event' is triggered, thus sparsifying control. Event-triggered control of multi-agent systems has been studied by many. \cite{dimarogonas2012distributed}, \cite{seyboth2011control}  prove consensus under a connected time-invariant graph for single integrator dynamics, while \cite{yu2015distributed}, \cite{zhu2014event} have extended the results for agents with general linear systems dynamics. \cite{chen2016event} has studied the consensus of time varying systems with non-linear dynamics under event-triggered control but under a constant interaction topology. Our work considers the broader case of time-varying graphs that can have different spanning-trees and we prove consensus of single-integrator systems under event-triggered control structure. 
\\
The paper is organized as follows. 
We brush up on graph theory and persistent excitation in section \ref{sec2}. The system dynamics for single integrator systems are introduced in section \ref{sec3}. In section \ref{sec4}, we introduce the notion of event-triggered control, define the trigger conditions and evaluate convergence under event-triggered control in section \ref{sec5}. We extend the results to switching graphs in section \ref{sec6} and show simulation results in section \ref{sec7}. The results that we obtained are summarized in section \ref{sec8}.
 
\section{Preliminaries}\label{sec2}
\subsection{Notions of graph theory}
In this work we consider agent interactions represented by undirected graphs $\mathcal{G}=\left(V, \, E\right)$, where $V=\left\lbrace v_1,v_2,\cdots v_n\right\rbrace$ denotes a  non-empty set of nodes and $ \left[ V\right]^{2} \supseteq E=\left\lbrace e_1,e_2,\cdots e_m\right\rbrace$ is the edge set, where $\left[ V\right]^{2}$ is the set of all subsets of $V$ containing two elements. Each node represents an agent of the system and each edge $\braces{v_i,v_j}$ signifies that the agents occupying nodes $v_i$ and $v_j$ can exchange information with each other. We define $D\braces{\mathcal{G}}\braces{=[d_{ij}]}\in \R^{n\times m}$ to be the incidence matrix associated with the graph $\mathcal{G}$ by arbitrarily assigning orientation to each edge $e_j\in E$. Then $[d_{ij}]=-1$ if  $v_i$ is the tail of $e_j$ ,$[d_{ij}]=1$ if  $v_i$ is the head of $e_j$,
$[d_{ij}]=0$ otherwise. The Laplacian of $\mathcal{G}$, 
\begin{align}
L\braces{\mathcal{G}}=D\braces{\mathcal{G}}D\braces{\mathcal{G}}^{\top}
\end{align}
is a symmetric square matrix that captures the inter-connections between each pairs of nodes. We model the time-varying interactions between the agents by assuming a constant, underlying graph $\mathcal{G}$ with edge weights that can be time-varying. The Laplacian can be tweaked to reflect the resultant time-varying graph $\tilde{\mathcal{G}}\braces{t}$ as, 
\begin{align}
L\braces{\tilde{\mathcal{G}}\braces{t}}=D\braces{\mathcal{G}} W\braces{t} D \braces{\mathcal{G}}^{\top}.
\end{align}
where $W\braces{t} \in \R^{m\times m}$ is a diagonal matrix which captures the time-varying nature of each interaction and $w_{ii}\geq 0\,\, \forall t$. 
We assume that the underlying graph $\mathcal{G}$ is connected and therefore contains a \textit{spanning-tree}, i.e. the edge set $E$ of $\mathcal{G}$ can be partitioned into two subsets $E=E_\tau \cup E_c$, where $E_{\tau}$ consists of the spanning-tree edges and $E_c$ contains the cycle edges. We also assume that the time-varying edge weight matrix $W\braces{t}$ is piece-wise continuous and satisfies the persistent excitation condition with constants $\braces{\mu_1, \mu_2, T}$.\\
\subsection{Persistence of excitation}
\begin{defin}\label{PE1}
~\cite[p.~72]{sastry2011adaptive} The signal $g(\cdot):\mathbb{R}^{\geq 0}\rightarrow \mathbb{R}^{n\times m}$ is \emph{Persistently Exciting} (PE) if there exist finite positive constants $\mu_1, \mu_2, T$ such that,
\begin{align}
\mu_2I_n\geq \int_t ^{t+T} g(\tau)g(\tau)^T d\tau\geq\mu_1I_n \hspace{1.5cm}  \forall t\geq t_0     \label{eq:pre1}
\end{align} 
\end{defin}
A function $g(\cdot)$ that satisfies the condition \eqref{eq:pre1} is said to be persistently exciting with constants $(\mu_1, \mu_2, T)$.\\
We say that a graph $\tilde{\mathcal{G}}$ is persistently exciting if its associated edge-weight matrix $W\braces{t}$ is persistently exciting.
\subsection{Other Conventions}
$\norm{\cdot}$ denotes the frobenius 2-norm on vectors and the induced 2-norm on matrices. $\lambda_{min} \braces{\cdot}$ and $\lambda_{min} \braces{\cdot}$ operate on square matrices and return the minimum and maximum eigenvalues respectively of the said matrix. Boldfaced $1$ and $0$ ($\mathbf{1},\mathbf{0}$) represent vectors with all ones and all zeroes respectively and $I$ is used to denote the identity matrix. Their dimensions can be inferred contextually if not mentioned explicitly. 

\section{Network Models}\label{sec3}
In this section we state the single  integrator dynamical equations and perform a series of linear transformations, to bring them to a form that we could work with later on. This section has been taken from \cite{chowdhury2016persistence} and presented here for reference of the readers. 
\subsection{Single Integrator}\label{Single integrator model}
Consider a multi-agent system with states $x_i \in \R$ for $i=1,2,3\cdots n$ with a connected underlying graph $\mathcal{G}$ and the Laplacian of the time-varying graphs $L\braces{\tilde{\mathcal{G}}\braces{t}}=[l_{ij}\braces{t}]$. The dynamics of each state with control $u_i\in \R$ can be written as,
\begin{align}
\dot{x}_{i}=&u_i \nonumber\\
u_i=&-k\sum_{j=1}^{n}l_{ij}(t)x_{j} 
\end{align}
for $t>t_0$ with initial condition $x_i\braces{t_0}\in\R$ for $i=1,2,3.....n$ and positive control gain $k\in \R$. The augmented dynamics of the system can be written in terms of the state vectors $x=\left[x_{1},x_{2},\cdots x_{n}\right]^{\top}\in\R^{n}$
as,
\begin{flalign}
\dot{x}(t)= & -kL\left(\tilde{\mathcal{G}}(t)\right)x \nonumber \\ 
= & -kD\left(\mathcal{G}\right)W(t)D\left(\mathcal{G}\right)^{\top}x\label{eq:Single integrator dynamics}
\end{flalign}
Taking cue from \cite{zelazo2011edge}, \citealp[p.~77-81]{mesbahi2010graph} we transform the consensus problem of equation \eqref{eq:Single integrator dynamics} into a stabilization problem by considering the edge states instead of node states. We effect the conversion through the following transformation,
\begin{flalign}
x_{e}= & D\left(\mathcal{G}\right)^{\top}x.\label{eq:edge transformation}
\end{flalign}
The dynamics of the edge states, on differentiation of (\ref{eq:edge transformation})
yields,
\begin{flalign}
\dot{x}_{e}= & -kL_{e}\left(\mathcal{G}\right)W\left(t\right)x_{e}\label{eq:1}
\end{flalign}
 where $L_e\left(\mathcal{G}\right)\in\R^{m\times m}$
represents the edge-Laplacian of the graph $\mathcal{G}$ and can be expressed as, $L_e\braces{\mathcal{G}}=D\braces{\mathcal{G}}^{\top}D\braces{\mathcal{G}}$
   as shown in \cite{zelazo14}. Also, using the fact
that the underlying graph contains a spanning-tree $\mathcal{G}_{\tau}$ we partition the edge states after a suitable permutation as follows,
\begin{align}
x_{e}= & \left[\begin{array}{c}
x_{\tau}\\
x_{c}
\end{array}\right]\label{eq:2}
\end{align}
for $x_{\tau} \in \R^{p}$ and $x_c \in \R^{m-p}$ where $p$ is the number of edges in $\mathcal{G}_\tau$. Similarly we can partition $D\braces{\mathcal{G}}$, $W\braces{t}$ as $D\left(\mathcal{G}\right)=  \left[D\left(\mathcal{G}_{\tau}\right)\,\,D\left(\mathcal{G}_{c}\right)\right]$, $
W\left(t\right)=  \left[\begin{array}{cc}
W_{\tau}\left(t\right) & 0\\
0 & W_{c}\left(t\right)
\end{array}\right]$. The edge-Laplacian, in terms of the partitions of $D\braces{\mathcal{G}}$ is then, 
\begin{flalign}
L_{e}\left(\mathcal{G}\right)= & \left[D\left(\mathcal{G}_{\tau}\right)\,\,D\left(\mathcal{G}_{c}\right)\right]^{\top}\left[D\left(\mathcal{G}_{\tau}\right)\,\,D\left(\mathcal{G}_{c}\right)\right]\nonumber\\
= & \left[\begin{array}{cc}
L_{e}\left(\mathcal{G}_{\tau}\right) & D\left(\mathcal{G}_{\tau}\right)^{\top}D\left(\mathcal{G}_{c}\right)\\
D\left(\mathcal{G}_{c}\right)^{\top}D\left(\mathcal{G}_{\tau}\right) & L_{e}\left(\mathcal{G}_{c}\right)
\end{array}\right].\label{eq:3}
\end{flalign}
We know that for connected graphs, $x_c$ can always be written as $x_c=Z^{\top}x_{\tau}$ as shown in \cite{sandhu2005cuts} where $Z=\left(L_e\braces{\mathcal{G}_{\tau}}\right)^{-1}D\left(\mathcal{G}_{\tau}\right)^{\top}D\left(\mathcal{G}_{c}\right)$. This is because the spanning-tree edges essentially capture the behaviour of all the edges. So we can focus our attention only on the spanning-tree edges. Using \eqref{eq:1}, \eqref{eq:2} and \eqref{eq:3}, we get $\dot{x}_{\tau}=  -kL_{e}\left(\mathcal{G}_{\tau}\right)RW\left(t\right)R^{\top}x_{\tau}$ where $R=\left[I_{p}\,\,Z\right]\in\R^{p\times m}$. The preceding transformations that helped us re-write the system dynamics in terms of $x_\tau$ are along the lines of \cite{zelazo2011edge} and \citealp[p.~77-81]{mesbahi2010graph}.  
 Since $L_{e}\left(\mathcal{G}_{\tau}\right)$ is symmetric and positive definite, it can be diagonalized as $L_{e}\left(\mathcal{G}_{\tau}\right)=\Gamma\Lambda\Gamma^{\top}$ for
some orthogonal matrix $\Gamma\in\R^{p\times p}$ and diagonal
matrix $\Lambda\in\R^{p\times p}$. Consider a change
of variable by the transformation $\Upsilon=\Gamma^{\top}x_{\tau}$. The above
equation becomes,
\begin{flalign}
\dot{\Upsilon}= & -k\Lambda M\left(t\right)\Upsilon\label{eq:Single integrator final}
\end{flalign}
where $M\left(t\right)=\Gamma^{\top}RW\left(t\right)R^{\top}\Gamma\in \R^{p\times p}$.
As stated earlier, the consensus problem of equation (\ref{eq:Single integrator dynamics}) is equivalent to stabilization problem
of equation (\ref{eq:Single integrator final}). For the sake of further reference, we denote the preceding set of transformations from $x$ to $\Upsilon$ by $\psi:=\Gamma^{\top}\left[I_p\; 0_{m-p}\right]D\braces{\mathcal{G}}^{\top}$. So we have, $\Upsilon=\psi x$. 
We state \cite[Theorem~5]{chowdhury2016persistence} and use the result to obtain the rate of convergence to consensus of a single integrator system defined by equations \eqref{eq:Single integrator dynamics}.
\begin{Thrm}\label{thm1}
(\cite[Theorem~5]{chowdhury2016persistence}) Consider the closed-loop consensus dynamics~(\ref{eq:Single integrator dynamics}). Assume that, the underlying graph $\mathcal{G}$ is connected. The states of the closed-loop dynamics $x(t)$ with time-varying communication topology characterized by $ W(t) $, achieve consensus exponentially, if there exists a spanning tree with corresponding edge-weight matrix $ W_{\tau}(t) $ that is persistently exciting. Further the convergence rate $ \alpha_v$ to consensus is bounded below by,
\begin{align*}
\alpha_v \geq {\dfrac{1}{2T} \ln \frac{1}{\left[ 1-\frac{2k\lambda_{\min}(\Lambda) \mu_1}{\left( 1+k\sqrt{p} \parallel \Lambda \parallel \mu_2 \right)^2}\right]  }}  
\end{align*}
where, $T, \mu_1$ and $\mu_2$ are the constants appearing in Definition~\ref{PE1} and $\Lambda$ is a diagonal matrix containing the eigenvalues of the spanning tree edge Laplacian matrix.
\end{Thrm}
Further,
\begin{align}
\left\Vert \Upsilon(t)\right\Vert \leq & m_{v}e^{-\alpha_{v}(t-t_{0})}\left\Vert \Upsilon(t_{0})\right\Vert \label{eq:stm inequality}
\end{align}
where $\alpha_v$ and $m_v$ can be calculated from the underlying graph and control gains.
\textbf{Note}:
The relationship between $\left\Vert x_e\right\Vert $ and $\norm{\Upsilon}$
can be established in the following way.
\begin{align}
\left\Vert x_{e}\right\Vert = & \sqrt{\left\Vert x_{\tau}\right\Vert ^{2}+\left\Vert x_{c}\right\Vert ^{2}}\nonumber\\
\leq & \left\Vert x_{\tau}\right\Vert \sqrt{1+\left\Vert Z^{\top}\right\Vert ^{2}}\nonumber
\end{align}
Using the fact that $\Gamma$ is orthogonal, $x_{\tau}=\Gamma \Upsilon$,
\begin{align}
\left\Vert x_e\right\Vert \leq & \rho\left\Vert \Upsilon\right\Vert\label{eq:ineq rel} 
\end{align}
where $\rho=\left\Vert \Gamma\right\Vert \sqrt{1+\left\Vert Z^{\top}\right\Vert ^{2}}$. The induced 2-norm of $Z$ can be calculated as $\lambda_{max}\left(Z^{\top}Z\right)^{\frac{1}{2}}$.

\section{Event-Triggered Control}\label{sec4}
In this section we outline the event-triggered control strategies and show how it modifies the single-integrator  dynamics defined by equation \eqref{eq:Single integrator dynamics}.
Under the event-triggered control paradigm each agent broadcasts a (piecewise)constant value, $\hat{x}_i$ which is updated to the current value of the state whenever an 'event is triggered'. The control applied by each agent would then be,
\begin{align}
u_i=&-k\sum_{j=1}^{n}l_{ij}(t)\hat{x}_{j} 
\end{align}
To define an event, we introduce error variables $e_{i}\left(t\right)=\hat{x}_{i}\left(t\right)-x_{i}\left(t\right)$ which denote the difference between the broadcasted value and the current value of the state  for each agent. The trigger condition that updates the broadcasted states $\hat{x}$ effectively shapes the behaviour of the system. We define the trigger condition using a trigger function for each state $f_i\braces{t,e_i}:\R\rightarrow\R$ . An event is said to be 'triggered' when $f_i>0$. Once an event is 'triggered' say at time $t^{*}$, the broadcast value is updated $\ie$ $\hat{x}_i\braces{t}=x_i\braces{t^{*}} \implies e_i\braces{t^{*}}=0$ for $t^*\leq t < t^{'}$ where $t^{'} $ is the time instant when the next subsequent event is triggered. 
We define two trigger functions as shown in \cite{seyboth2011control}- the static trigger and the dynamic trigger.
\begin{enumerate}
\item \textbf{Static Trigger Function} 
\begin{align}
f_{i}(e_{i}(t))= & \left\Vert e_{i}(t)\right\Vert-c\label{static trigger}
\end{align}
\item \textbf{Dynamic Trigger Function}
\begin{align}
f_{i}(t,\, e_{i}(t))= & \left\Vert e_{i}(t)\right\Vert-ce^{-\beta\braces t}\label{Dynamic Trigger}
\end{align}
where $c > 0$  
\end{enumerate}
The static trigger function can be seen as a special case of the dynamic trigger function with $\beta=0$. So we will perform our analysis with \eqref{Dynamic Trigger} as the trigger function, and substitute $\beta=0$ when we want to evaluate the static-trigger case.
The dynamics of single integrator systems under event-triggered control can be written as, 
 \begin{flalign}
\dot{x}(t)= & -kL\left(\tilde{\mathcal{G}}\braces{t}\right)\braces{x+e}\label{eq:Single integrator-event}
\end{flalign} 
 where $e=[e_1,e_2\cdots e_n]^{\top}\in \R^n$.
The equivalent stabilization problem under event triggered control for single integrator will then be,
\begin{align}
\dot{\Upsilon}= & -k\Lambda M\left(t\right)\left(\Upsilon+\tilde{e}\right)\label{eq:Single integrator -event2}
\end{align}
where $\tilde{e}=\psi e$.\\
\vspace{-5pt}
\subsection{Bounds on the error variables}
In this section, we obtain bounds on $\tilde{e}$ in terms of the bounds on $e$. The trigger function is so designed that each $e_i$ is always upper-bounded. We can see that,
\begin{align}
\norm{e_i}\leq & ce^{-\beta t}\label{eq:4}
\end{align}
We can relate the bounds on $e$ to bounds on $\tilde{e}$ in the following way. We know that $\tilde{e}=\psi\, e$. Let us define $\bar{e}:=D\braces{\mathcal{G}}^{\top}e$. From the structure of $D\braces{\mathcal{G}}$, we get
\begin{align}
\bar{e}_i=e_j-e_k\label{eq:9}
\end{align} 
for $i=1,2,\cdots m$ and $j,k$ chosen on the basis of $D\braces{\mathcal{G}}$. Using \eqref{eq:4},  \eqref{eq:9} can be rewritten as,
\begin{align}
\norm{\bar{e}_i}&=\norm{e_j-e_k}\nonumber\\
&\leq \norm{e_j}+\norm{e_k}\nonumber\\
&\leq 2ce^{-\beta t}.\label{eq:10}
\end{align}
Also,
\begin{align}
\tilde{e}&=\psi e\nonumber\\
&=\Gamma^{\top}\left[I_p\; 0_{m-p}\right]\bar{e}\nonumber\\
&=\Gamma^{\top}\left[\bar{e}_{1}\,\bar{e}_{2}\,\cdots \bar{e}_{p}\right]^{\top}.
\end{align}
 Using the bound on each $\bar{e}_{i}$ from \eqref{eq:10},
\begin{align}
\norm{\tilde{e}}&\leq\norm{\Gamma}\;\norm{\left[\bar{e}_{1}\,\bar{e}_{2}\,\cdots \bar{e}_{p}\right]}\nonumber \\
&\leq\sqrt{p} \norm{\Gamma}\norm{\bar{e}_i}\nonumber\\
&\leq 2c\sqrt{p} \norm{\Gamma}e^{-\beta t}.
\end{align}
Defining  $C:=2c\sqrt{p}\left\Vert\Gamma \right\Vert $, we can write the above inequality to be,
\begin{align}
\left\Vert\tilde{e}(t)\right\Vert\leq Ce^{-\beta t}.\label{eq;error inequality single}
\end{align}

\section{Consensus Analysis}\label{sec5}
 
\begin{Thrm}\label{Single integrator theorem}
Consider a multi-agent system with single integrator dynamics as defined by equation \eqref{eq:Single integrator dynamics}. Assume that the underlying graph ($\mathcal{G}$), representing the interaction between the agents be connected, with $p$ edges in the spanning-tree. If the spanning tree edge weight matrix $W_{\tau}\braces{t}$ is persistently exciting with constants ($\mu_1$,$\mu_2$,$T$), then
\begin{enumerate}
\item on application of event-triggered control with a static trigger function defined by equation \eqref{static trigger}, the edge states $x_e$ of the system exponentially converge to a ball around the origin defined by $\norm{x_e}\leq \rho \kappa_2^{m}$.
\item on application of event-triggered control with a dynamic trigger function defined by equation \eqref{Dynamic Trigger}, the edge states $x_e$ of the system exponentially converge to origin. Also the rate of convergence is lower bounded by $\beta$ as defined in \eqref{Dynamic Trigger}.
\end{enumerate}
where, $\kappa_{2}=\frac{k\left\Vert \Lambda\right\Vert m_{v}Ce^{\alpha_{v}t_{0}+2\alpha_vT}\mu_2}{e^{\alpha_{v}T}-1}$ and $\rho$ is defined as in \eqref{eq:ineq rel}. Also the closed loop systems in the cases of static and dynamic triggers does not exhibit zeno behaviour when $\beta$ is chosen to be greater than $\alpha_v$. \end{Thrm}
\textbf{Note}: When a connected graph has multiple spanning-trees, $\kappa_2^{M}$ and $\kappa_2^{m}$ are the largest and smallest values of $\kappa_2$ that can be calculated considering each different spanning-tree.
\begin{pf}
Let $\phi\braces{t,t_0}$ be the state transition matrix corresponding to system defined by equation \eqref{eq:Single integrator final} . The solution of the system can be written using $\phi\braces{t,t_0}$ as,
\begin{align*}
\Upsilon(t)= & \phi(t,t_{0})\Upsilon(t_{0})
\end{align*}
We obtain a bound on $\norm{\phi(t,t_{0})}$ by the following steps.
\begin{align}
\norm{\Upsilon(t)}= & \norm{\phi(t,t_{0})\Upsilon(t_{0})}\nonumber\\
= & \frac{\norm{\phi(t,t_{0})\Upsilon(t_{0})}}{\norm{\Upsilon(t_{0})}}\norm{\Upsilon(t_{0})}\label{eq:12}
\end{align} 
Comparing \eqref{eq:12} and \eqref{eq:stm inequality} we can conclude that,
\begin{align}
\frac{\norm{\phi(t,t_{0})\Upsilon(t_{0})}}{\norm{\Upsilon(t_{0})}}&\leq m_{v}e^{-\alpha_{v}(t-t_{0})}\nonumber\\
\underset{\Upsilon(t_{0})}{sup}\frac{\norm{\phi(t,t_{0})\Upsilon(t_{0})}}{\norm{\Upsilon(t_{0})}}=\norm{\phi(t,t_{0})}&\leq m_{v}e^{-\alpha_{v}(t-t_{0})}.\label{eq:stm inequality2}
\end{align}
The last statement holds true because $\Upsilon\braces{t_0}$ can be arbitrary. Consider the single integrator multi-agent system with event-triggered control defined by the equation \eqref{eq:Single integrator -event2}. The solution of this system can be expressed as ,
\begin{align}
\Upsilon(t)= & \phi(t,t_{0})\Upsilon(t_{0})-k\int_{t_{0}}^{t}\phi(t,\tau)\Lambda M(\tau)\tilde{e}(\tau)d\tau
\end{align}
We get the following inequality from the above equation,
\begin{align*}
\left\Vert \Upsilon(t)\right\Vert \leq & \left\Vert \phi(t,t_{0})\right\Vert \left\Vert \Upsilon(t_{0})\right\Vert +\\&k\left\Vert \Lambda\right\Vert\int_{t_{0}}^{t}\left\Vert \phi(t,\tau)\right\Vert \left\Vert M(\tau)\right\Vert \left\Vert \tilde{e}(\tau)\right\Vert d\tau
\end{align*}
Using the bounds on $\phi$ from equation \eqref{eq:stm inequality2}  and $\tilde{e}\braces{\tau}$ from equation \eqref{eq;error inequality single} we get,
\begin{align}
\left\Vert \Upsilon(t)\right\Vert \leq & m_{v}e^{-\alpha_{v}(t-t_{0})}\left\Vert \Upsilon(t_{0})\right\Vert +\nonumber\\&k\left\Vert \Lambda\right\Vert m_{v}Ce^{-\alpha_{v}t}\int_{t_{0}}^{t}e^{-(\beta-\alpha_{v})\tau}\left\Vert M(\tau)\right\Vert d\tau\label{eq:6}
\end{align}
To obtain a bound on the integral in the above inequality, we divide the interval $\left[t_0,t\right]$ into partitions of size $T$ i.e. $\left[t_0,t_0+T\right]$, $\left[t_0+T,t_0+2T\right]$ and so on . The number of such partitions of length $T$ possible will be given by $\theta=\left\lfloor \frac{t-t_{0}}{T}\right\rfloor $, where $\left\lfloor \cdot\right\rfloor$ is the floor function . The last partition can then be written as $\left[t_0+\theta T,t\right]$. The integral in \eqref{eq:6} can then be written as,$\int_{t_{0}}^{t_{0}+T}e^{-(\beta-\alpha_{v})\tau}\left\Vert M(\tau)\right\Vert d\tau+ \int_{t_{0}+T}^{t_{0}+2T}e^{-(\beta-\alpha_{v})\tau}\left\Vert M(\tau)\right\Vert d\tau \cdots +\int_{t_{0}+\theta T}^{t}e^{-(\beta-\alpha_{v})\tau}\left\Vert M(\tau)\right\Vert d\tau$. Applying H$\ddot{o}$lder inequality with $p=\infty$ and $q=1$ and using the fact that $M\braces{\tau}$ is persistently exciting and $\beta < \alpha_{v}$ we can obtain the following bound on each interval,
\begin{align*}
\int_{t'}^{t'+T}e^{-(\beta-\alpha_{v})\tau}\left\Vert M(\tau)\right\Vert d\tau\leq & \mu_{2}sup_{\tau\in[t',t'+T]}e^{-(\beta-\alpha_{v})\tau} \\
\leq & \mu_{2}e^{-(\beta-\alpha_{v})(t'+T)} 
\end{align*}

Using this upper bound  on each integral we get\begin{align*}
 \int_{t_0}^{t}e^{-(\beta-\alpha_{v})\tau}\left\Vert M(\tau)\right\Vert d\tau \leq \mu_{2}e^{-(\beta-\alpha_{v})T}\left(e^{-(\beta-\alpha_{v})t_0}+\right. \\ e^{-(\beta-\alpha_{v})(t_0+T)}+\cdots e^{-(\beta-\alpha_{v})(t_0+\theta T)}\left.\right)
\end{align*}
Using the property of the sum of terms in a geometric progression, we get
\begin{align}
 \int_{t_0}^{t}e^{-(\beta-\alpha_{v})\tau}\left\Vert M(\tau)\right\Vert &d\tau \leq\nonumber\\&\mu_{2}e^{-(\beta-\alpha_{v})\braces{t_0+T}}\left(\frac{1-e^{-(\beta-\alpha_{v})\theta T}}{1-e^{-(\beta-\alpha_{v})T}}\right)\label{eq:7}
\end{align}
Using inequality \eqref{eq:7} in inequality \eqref{eq:6},
\begin{align}
\left\Vert \Upsilon(t)\right\Vert &\leq m_{v}e^{-\alpha_{v}\braces{t-t_{0}}}\left\Vert \Upsilon(t_{0})\right\Vert+ \nonumber\\&k\left\Vert \Lambda\right\Vert m_{v}Ce^{-(\beta-\alpha_{v})(t_{0}+T)}\mu_2e^{-\alpha_{v}t}\left(\frac{1-e^{-(\beta-\alpha_{v})\theta T}}{1-e^{-(\beta-\alpha_{v})T}}\right)\label{eq:8}
\end{align}
We define $\kappa_{1}=e^{\alpha_{v}t_0}m_{v}\left\Vert \Upsilon(t_0)\right\Vert-\kappa_{3}$, $\kappa_2=e^{\braces{\alpha_v+\beta}T}\kappa_{3}$, 
\begin{align*}
\kappa_{3}=&\frac{k\left\Vert \Lambda\right\Vert m_{v}Ce^{-(\beta-\alpha_{v})(t_{0}+T)}\mu_2}{e^{-(\beta-\alpha_{v})T}-1}
\end{align*}
to be able to express \eqref{eq:8} as,
\begin{align}
\left\Vert \Upsilon(t)\right\Vert & \leq  \kappa_{1}e^{-\alpha_{v}t}+\kappa_{3}e^{-\beta \theta T-\alpha_{v}(\theta T-t)}
\end{align}
Using the properties of floor function, we can simplify the above inequality further,
\begin{align}
\left\Vert \Upsilon(t)\right\Vert & \leq \kappa_{1}e^{-\alpha_{v}t}+\kappa_{3}e^{-\beta \braces{\theta T-t}-\alpha_{v}(\theta T-t)-\beta t}\nonumber\nextline
\leq \kappa_{1}e^{-\alpha_{v}t}+\kappa_{2}e^{-\beta t}\label{eq:single int inequality y}\\
\implies \norm{x_e(t)}& \leq \rho \braces{ \kappa_{1}e^{-\alpha_{v}t}+\kappa_{2}e^{-\beta t}}.\label{eq:single int inequality x}
\end{align}
From the above expression it can clearly be seen that for the dynamic trigger case, $lim_{t\rightarrow\infty}\left\Vert x_e(t)\right\Vert =0$. This guarantees consensus of the states $x$. Also as $\beta < \alpha_v$ by choice, the rate of decay of the RHS of \eqref{eq:single int inequality x} will be dominated by $\beta$.
For the static trigger case, $\beta=0$ 
\begin{align*}
\left\Vert x_e(t)\right\Vert & \leq \rho \braces{\kappa_{1}e^{-\alpha_{v}t}+\kappa_{2}}\\
lim_{t\rightarrow\infty}\left\Vert x_e(t)\right\Vert  & \leq lim_{t\rightarrow\infty}\rho\kappa_{1}e^{-\alpha_{v}t}+lim_{t\rightarrow\infty}\rho\kappa_{2}\nextline
\leq \rho\kappa_{2}
\end{align*}
It can be seen that the edge-states converge to a ball around the origin $\norm{x_e}\leq \rho\kappa_{2}$ exponentially with a rate of convergence bounded below by $\alpha_v$. \vspace{3pt}\\
\textbf{Ruling out zeno behaviour in closed loop systems}\\
The system states $x_i$ and error states $e_i$ together form a hybrid system, which makes it necessary for us to ensure that zeno behaviour does not occur. Say that for the $i^{th}$ agent, an event is triggered at a time instant $t_1$ and another consecutive event is triggered at time $t_2$ for some $t_0\leq t_1<t_2$. To rule out the occurrence of zeno behaviour, it is sufficient to show that there exists a positive, non-zero lower bound on $\gamma:=t_2-t_1$. We have $\dot{e}_i=-\dot{x}_i$. Consider the following set of inequalities for time $t_1\leq t < t_2$.
\begin{align*}
\norm{\dot{e}_i}&=\norm{\dot{x}_i}\leq \norm{\dot{x}}=\norm{kD\left(\mathcal{G}\right)W(t)D\left(\mathcal{G}\right)^{\top}x\braces{t_1}}\\
&\leq\, k\norm{D\braces{\mathcal{G}}}\,\norm{W(t)}\,\norm{x_e\braces{t_1}}
\end{align*}
Using \eqref{eq:single int inequality x} we can rewrite the above inequality as,
\begin{align}
\norm{\dot{e}_i}&\leq k\rho\norm{D\braces{\mathcal{G}}}\,\norm{W(t)}\,\braces{ \kappa_{1}e^{-\alpha_{v}t_1}+\kappa_{2}e^{-\beta t_1}}.\label{eq:13}
\end{align}
Also,
\begin{align}
\int_{t_1}^{t_2}\norm{\dot{e}_i}dt&\geq \Big\Vert\int_{t_1}^{t_2}\dot{e}_i\, dt\,\,\Big\Vert=\norm{e_i\braces{t_2}}.\label{eq:14}
\end{align}
We substitute $\norm{e_i\braces{t_2}}=ce^{-\beta t_2}$ as an event is triggered at $t_2$.
Integrating the inequality \eqref{eq:13} with limits $t_1$ and $t_2$ and using the fact that the edge weights are  bounded such that $\norm{W(t)}\leq \omega$ and \eqref{eq:14} we get,
\begin{align*}
\norm{e_i\braces{t_2}}&\leq \int_{t_1}^{t_2} k\rho\omega\norm{D\braces{\mathcal{G}}}\,\braces{ \kappa_{1}e^{-\alpha_{v}t_1}+\kappa_{2}e^{-\beta t_1}}dt\\
ce^{-\beta t_2}&\leq k\rho\omega\norm{D\braces{\mathcal{G}}}\,\braces{ \kappa_{1}e^{-\alpha_{v}t_1}+\kappa_{2}e^{-\beta t_1}}\gamma \\
ce^{-\beta \gamma}&\leq k\rho\omega\norm{D\braces{\mathcal{G}}}\,\braces{ \kappa_{1}e^{-(\alpha_{v}-\beta)t_1}+\kappa_{2}}\gamma
\end{align*}
Rearranging the above inequality we get,
\begin{align}
\gamma\, e^{\beta \gamma} &\geq \frac{c}{k\rho\omega\norm{D\braces{\mathcal{G}}}\,\braces{ \kappa_{1}e^{-(\alpha_{v}-\beta)t_1}+\kappa_{2}}}\label{eq:zeno inequality 1} 
\end{align}
It is evident that $\gamma>0$ as the RHS of \eqref{eq:zeno inequality 1} is strictly positive.
The minimum value of $\gamma$ is a solution of the following equation,
\begin{align}
\gamma\, e^{\beta \gamma} &= \frac{c}{k\rho\omega\norm{D\braces{\mathcal{G}}}\,\braces{ \kappa_{1}+\kappa_{2}}}\label{eq:zeno dynamic upper bound} 
\end{align}
This proves that zeno behaviour does not occur in the dynanic trigger case. When $\beta=0$, the lower bound on gamma is  the RHS of equation \eqref{eq:zeno dynamic upper bound}, which rules out zeno behaviour in static trigger case as well. The preceding arguments on ruling out zeno behaviour are similar to those provided in \cite{seyboth2011control} for time-invariant graphs. 
\end{pf}
%
\subsection{Consensus in Switching Topologies}\label{sec6}
Theorem \ref{Single integrator theorem} is not valid for switching topologies because our spanning tree and thereby the edge states $x_e$ itself can be changing. The following corollary extends the aforementioned theorem to agents interacting via switching topologies.
\begin{Corr} \label{main4}
Let $ t_1, t_2,\cdots $ be the infinite time sequence of graph switching instants with, $ t_{i+1}-t_i \geq t_L $ for some positive $ t_L $, and $ i=0,1, \cdots $. Consider the agent dynamics (\ref{eq:Single integrator-event}) corresponding to a single integrator system with event-triggered control. If there exists an infinite sequence of contiguous, non-empty and uniformly bounded time intervals $\tau\braces{j,1}$ $\braces{Let \;\tau\braces{j,l}=\left[ t_{i_j}, t_{i_{j+l}}\right)}$; $ j=1,2,\cdots $ starting at $ t_{i_{1}}=t_0 $, with the property that the union of the undirected graphs across each such interval has a spanning tree, then
\begin{enumerate}
\item with static-trigger, the edge-state $x_e$ converges to a ball of radius $\norm{x_e}\leq \rho\kappa_2^{M}$.
\item with dynamic-trigger, the edge-state $x_e$ converges exponentially to origin at rate greater than $\beta$.
\end{enumerate}
\end{Corr}
\begin{pf}
The case of switching topologies differs from the setting of theorem \ref{Single integrator theorem} by the fact that the union of graphs in each time interval $\tau\braces{j,1}$ contains a different spanning-tree compared to the occurrence of the same spanning tree in theorem \ref{Single integrator theorem}. We show that Corollary \ref{main4} can be treated as an extension of Theorem \ref{Single integrator theorem} using the results of \emph{Van der Waerden's theorem} (\cite[p.~29]{graham1980ramsey}). The collection of possible spanning-tress forms a finite, non-empty set. Taking each possible spanning-tree to be a different colour, by Theorem \cite[p.~29]{graham1980ramsey} it is possible to find $N$ such that in the interval $\braces{t_{i_1},t_{i_{N}}}$, the same spanning-tree occurs in each $\tau\braces{1+(j-1)d,1}$ for $j=1,2,\cdots k$ and some $d>0$. This allows us to select a persistence window $T>(d+1)t_{max}$ where $t_{max}=max_j\; \tau\braces{j,1}$ for $j=1,2,\cdots N$. With the aforementioned selection of $T$, Theorem \ref{Single integrator theorem} can be invoked to prove the convergence of the edge states $x_e$ to a ball around origin in the case of static trigger and to origin in the case of dynamic trigger. As we cannot predict which particular spanning-tree repeats in each interval $\tau\braces{1+(i-1)d,1}$, the least conservative bounds are chosen.
\end{pf}
 
\section{Simulations}\label{sec7}
\begin{figure}[h!]
\centering
	\includegraphics[scale=1.10]{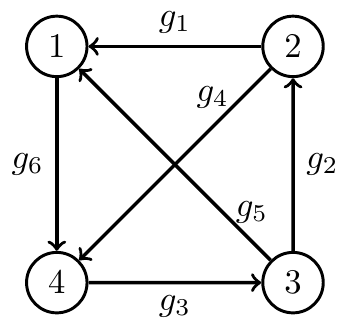}
\caption{\small{Underlying graph of arbitrary orientation}}
\label{fig1}
\end{figure}

\begin{figure}[h!]
\centering
	\includegraphics[scale=0.4]{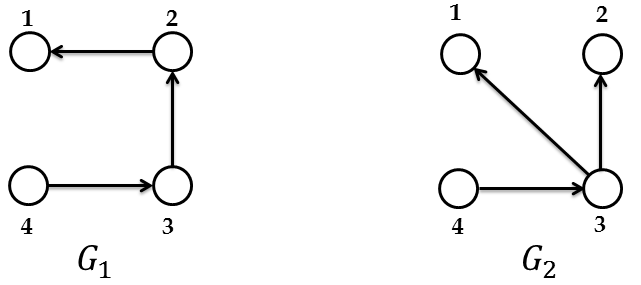}
\caption{\small{Spanning Trees Considered}}
\label{fig2}
\end{figure}

A multi-agent system with four agents under switching graphs was simulated using Matlab$^ \copyright$ for the static and dynamic trigger cases. The underlying communication topology considered is shown in figure \ref{fig1} and the different spanning-trees that were switched between are shown in figure \ref{fig2}.

\begin{figure}[h!]
\centering
	\includegraphics[scale=0.7]{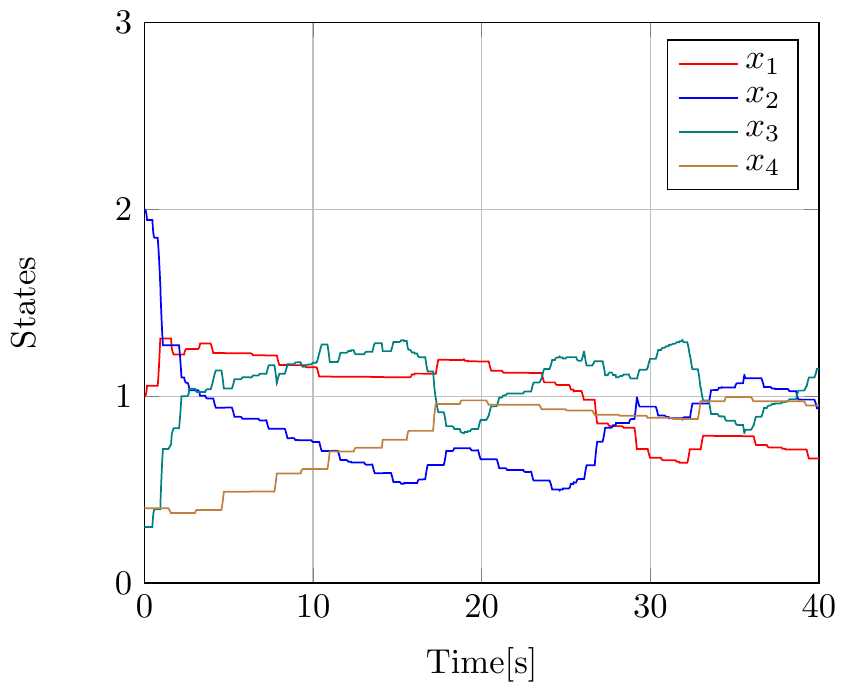}
\caption{\small{Evolution of states under static trigger with the graph switching between spanning trees $\mathcal{G}_1$ and $\mathcal{G}_2$ in contiguous intervals}}
\label{fig3}
\end{figure}

\begin{figure}[h!]
\centering
	\includegraphics[scale=0.7]{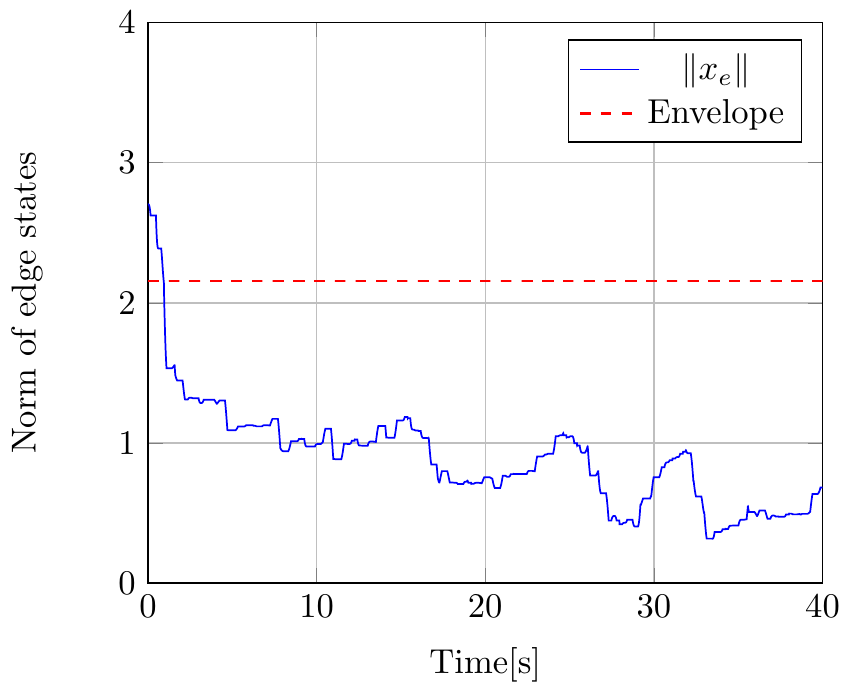}
\caption{\small{Evolution of norm of edge-states under static trigger with the graph switching between spanning trees $\mathcal{G}_1$ and $\mathcal{G}_2$ in contiguous intervals}}
\label{fig4}
\end{figure}

\begin{figure}[h!]
\centering
	\includegraphics[scale=0.7]{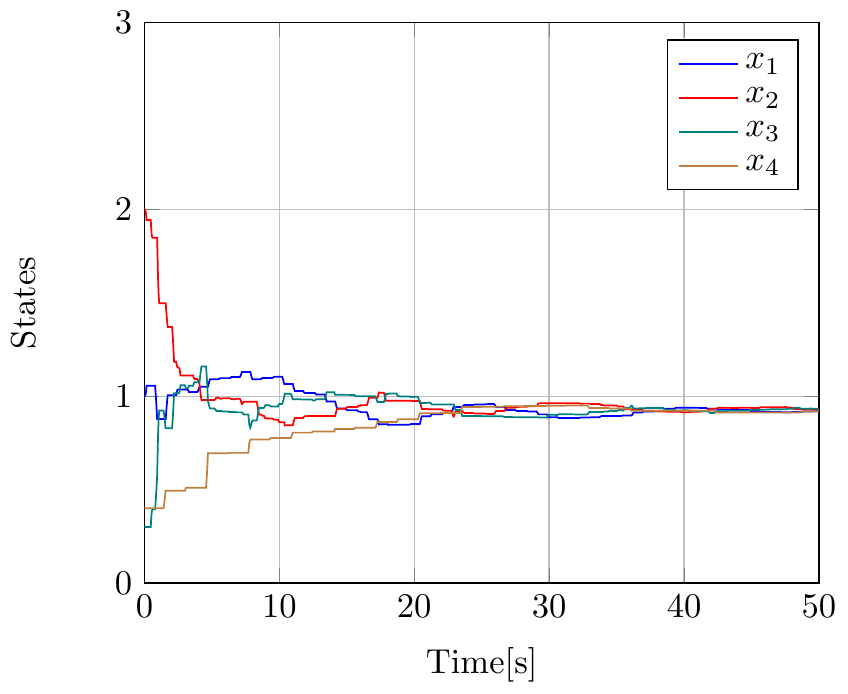}
\caption{\small{Evolution of states under dynamic trigger with the graph switching between spanning trees $\mathcal{G}_1$ and $\mathcal{G}_2$ in contiguous intervals}}
\label{fig5}
\end{figure}

\begin{figure}[h!]
\centering
	\includegraphics[scale=0.7]{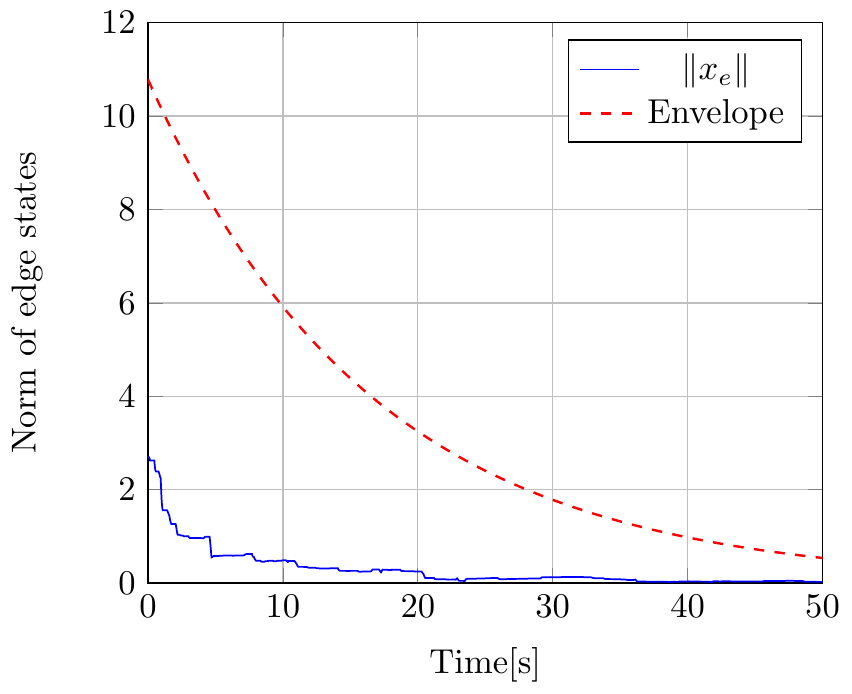}
\caption{\small{Evolution of norm of edge-states under dynamic trigger with the graph switching between spanning trees $\mathcal{G}_1$ and $\mathcal{G}_2$ in contiguous intervals}}
\label{fig6}
\end{figure}

The edge-weights were chosen as $g_i$ = square$(4*t,20-(i-1)0.1\pi)+1).*sin(5*t)$ for $i=1,2\cdots 4$ and $g_6=0$, where the function square$\braces{at,b}$ for $a,b\in\R$ generates a square wave of unit amplitude, period $\frac{2\pi}{a}$ and duty-cycle $\braces{\frac{T_{\text{on}}}{T_{\text{off}}+T_{\text{on}}}}$ $b$. The aforementioned $g_i$ are defined to emulate real scenarios where there might be instances when no edges are active. The initial value of the states was taken to be $x_{0}=\left[1\: 2\: 0.3\: 0.4\right] ^{\top}$ Plots \ref{fig3} and \ref{fig4} show the evolution of the system states $x$ and the norm of the edge states $\norm{x_e}$ under the static trigger with $c=0.5$ (refer equation \eqref{static trigger}). The evolution of the states with dynamic trigger function with $\beta=0.06$ and $c=0.5$ (refer equation \eqref{Dynamic Trigger}) is plotted in figures (\ref{fig5}) and (\ref{fig6}). The bounds were calculated using inequalities presented in Corollary \ref{main4} and plotted along with the norm of the edge states. These plots show the convergence of the edge states $x_e$ to a ball around the origin in case of static trigger and consensus of states $x$ in the case of dynamic trigger.
   \section{Conclusions}\label{sec8}
The application of event-triggered control to classical consensus algorithms with time-varying, persistently exciting topologies guarantees consensus with dynamic trigger function. Under the more practically implementable static trigger function, the edge-states converge to a ball around the origin. For switching topologies, we utilize the work of \cite{chowdhury2016persistence} to show that we can extend the results of the persistent, continuously varying graphs to the case of switching topologies. The convergence bounds thus obtained depend on the 'slowest' spanning tree.  
\small
\bibliography{bib}             
\end{document}